\documentclass{ws-procs975x65}

\newcommand{\apj}{\mbox{\it Astrophys. J.}}
\newcommand{\apjl}{\mbox{\it Astrophys. J.}}
\newcommand{\apjs}{\mbox{\it Astrophys. J.}}
\newcommand{\aap}{\mbox{\it Astron. Astrophys.}}
\newcommand{\araa}{\mbox{\it Annu. Rev. Astron. Astrophys.}}
\newcommand{\apss}{\mbox{\it Astrophys. ans Space Science}}

\newcommand{\jcap}{\mbox{\it J. Cosmol. Astropart. Phys.}}
\newcommand{\mnras}{\mbox{\it Mon. Not. R. Astron. Soc.}}
\newcommand{\nat}{\mbox{\it Nature}}

\newcommand{\physrep}{\mbox{\it Phys. Rep.}}

\newcommand{\keV}{\rm{\, keV }}

\newcommand{\beq}{\begin{equation}}
\newcommand{\eeq}{\end{equation}}
\newcommand{\ba}{\begin{array}}
\newcommand{\ea}{\end{array}}

\def\gtsima{$\; \buildrel > \over \sim \;$}
\def\ltsima{$\; \buildrel < \over \sim \;$}

\def\gsim{\lower.5ex\hbox{\gtsima}}
\def\lsim{\lower.5ex\hbox{\ltsima}}

\begin{document}

\title{Photospheric Emission in Gamma-Ray Bursts}

\author{Asaf Pe'er$^*$}

\address{Physics Department, University College Cork,\\
Cork, Ireland\\
$^*$E-mail: a.peer@ucc.ie}

\author{Felix Ryde}

\address{Department of Physics\\
KTH Royal Institute of Technology, and the Oskar Klein Centre\\
AlbaNova, SE-106 91 Stockholm, Sweden}

\begin{abstract}
A major breakthrough in our understanding of gamma-ray bursts (GRB)
prompt emission physics occurred in the last few years, with the
realization that a thermal component accompanies the over-all
non-thermal prompt spectra. This thermal part is important by itself,
as it provides direct probe of the physics in the innermost outflow
regions. It further has an indirect importance, as a source of seed
photons for inverse-Compton scattering, thereby it contributs to the
non-thermal part as well. In this short review, we highlight some key
recent developments. Observationally, although so far it was clearly
identified only in a minority of bursts, there are indirect evidence
that thermal component exists in a very large fraction of GRBs,
possibly close to 100\%. Theoretically, the existence of thermal
component have a large number of implications as a probe of underlying
GRB physics. Some surprising implications include its use as a probe
of the jet dynamics, geometry and magnetization.
\end{abstract}

\keywords{Gamma-rays: bursts; hydrodynamics; radiation mechanism: non-thermal; radiation mechanism: thermal}

\bodymatter

\section{Introduction}
\label{sec:intro}

Our understanding of gamma-ray bursts (GRB) prompt emission have been
revolutionized in the past few years. This is thanks to new
observatories, in particular the {\it Swift} and {\it Fermi}
satellites, new methods of data analysis, and new theoretical ideas of
interpreting these results. As a result of these progresses, we think
that it is fair to claim that we reached a point in time where we are
witnessing a 'paradigm shift' in our understanding of this
phenomena. A major part of this paradigm shift is attributed to the
realization that a thermal component exists in a large fraction of GRBs,
and the realization of its importance as providing a new tool to study
the underlying GRB physics.

This rapid progress in this field manifests itself in a large number
of excellent reviews that were written in the last few years. A
partial list of reviews that were published only in the last five
years include reviews by references \cite{GM12, Bucciantini12, GR13, Daigne13, Vereschagin14,
  Zhang14, Berger14, MR14, KZ15, Peer15}. This short review is not
aimed at competing with any of the above, but rather to highlight one
aspect of the progress, which we find of particular importance: that
of thermal emission component that is observed during the prompt phase
of many GRBs. There are currently good reasons to believe that such a
component in fact exists in many GRBs in which it is not directly
observed, the reason for this being its distortion by various
processes. As we will discuss here, this component, whether
independently as well as in combination with other, non-thermal parts of the
spectra, provides a wealth of novel ways of interpreting and
understanding the data. It can be used to study underlying physical
properties of the GRB outflow such as the jet geometry or jet
magnetization, that do not seem, at first sight, to be related to thermal
emission. As a consequence, it may very well hold the key to a more
complete understanding of the underlying physics of the chain of
events that eventually results in the production of a GRB.

We begin this review by a short historical overview in section
\ref{sec:2}. We then discuss the current observational status in
section \ref{sec:3}, and present theoretical ideas in section \ref{sec:4}. We
summarize and conclude in section \ref{sec:5}.

\section{Historical overview}
\label{sec:2}

Interestingly, the very first works in which it was realized that GRBs
are of cosmological origin, proposed that the emission should be
(quasi)-thermal \cite{Goodman86, Pac86, Pac90, SP90, RM92,
  Thompson94}. This idea originates from the realization that the huge
amount of energy, $\gsim 10^{53}$~erg released in a small volume, of
typical size $r \sim 10^7 - 10^8$~cm (that is deduced from light
crossing time arguments and the rapid, $\gsim$~ms variability
observed) must result in an extremely high optical depth to scattering
by particles in the plasma. The rapid interactions between the
energetic photons and particles / low energy photons (in producing
pairs) lead to the formation of a ``fireball'', similar in nature
to the early evolution of the expanding universe \cite{CR78}. In this
regime of optical depth $\tau \gg 1$, all emerging radiation must be thermal. The
observed spectrum, though, was predicted to be somewhat distorted from
a pure ``Planck'' function, due to light aberration effects in the
relativistically expanding winds (see Ref. \refcite{Goodman86}).

These ideas could be tested in the early 1990's, following the launch
of the {\it Compton gamma ray observatory (CGRO)} in 1991 that led to
the accumulation of detailed spectral data. Contrary to the initial
expectations, {\it CGRO} spectral data was found to be inconsistent
with the initial predictions. Data accumulated mainly by the BATSE
instrument on board the {\it CGRO} showed clearly that the observed
spectral shape of the vast majority of GRBs do not resemble a
``Planck'' function. Rather, the (time integrated) spectra could be
easily fitted with a nearly featureless broken power law spectra,
which peaks at the sub MeV energy range. This became known is as the
``Band'' function (after the late David Band; \cite{Band+93,
  Preece+00, Kaneko+06, Kaneko+08, Goldstein+13}).

The leading theoretical interpretation of this non-thermal spectra
was, and still is, synchrotron emission by relativistic electrons
\cite{MLR93, Usov94, Tavani96, Cohen+97, Schaefer+98, CD99, Frontera+00,
  BB04}. This is a very common mechanism that is capable of explaining
non-thermal emission in many different astronomical objects, from
solar flares to active galaxies. As such, it is well studied since the
1960's \cite{GS65}, and its basic theoretical framework appears in
many textbooks [e.g., \cite{RL79}]. Further support of this idea came
from the fact that synchrotron emission fits very well GRB afterglow
emission (at least during late times; see \cite{WRM97} onward). The
synchrotron emission (presumably peaking at the sub MeV range to match
the prompt emission data) is expected to be accompanied by
inverse-Compton scattering at higher energies (synchrotron self-Compton;
SSC). Alternative models suggested that energetic protons may have a
substantial contribution to the spectra via proton-synchrotron
emission or photo-pair production \cite{Totani98, CD99}. The hadronic
models, though typically require the deposition of a very large amount
of kinetic energy, due to the much less efficient emission from
protons as compared to electrons.

Although the synchrotron emission model became widely accepted by the mid
1990's, already in the late 1990's evidence began to accumulate that
the low energy spectral slopes (below the sub MeV peak) observed in
the vast majority of GRBs are steeper than allowed by the synchrotron
or synchrotron-SSC models \cite{Crider+97, Preece+98, preece+02,
  GCG03}. A second difficulty is the inability of this model to
explain the observed correlation between peak energy and luminosity
\cite{Golenetskii+83, Amati+02} without invoking additional
assumptions \cite{ZM02, LZ04}.

From a theoretical perspective, the synchrotron model relies on the
existence of energetic particles and strong magnetic fields. Within
the context of the internal shock model, particles gain their energy
following internal shocks that dissipate the outflow kinetic
energy. However, a well known problem is the very low efficiency in
energy conversion, typically no more than a few \% \cite{MMM95,
  Kobayashi+97, PSM99, LGC99, Kumar99, Spada+00, GSE01, MZ09}. A
second theoretical problem that arises from fitting the data is that
the required values of the magnetic field needed to produce the sub MeV
peak are close to equipartition, while fits of the afterglow show that
the magnetic field produces at shock fronts is typically two orders of
magnitude below equipartition, and in many cases less
\cite{WRM97,SBK14}. A third problem is the fact that the energy of the
peak is very sensitive to the model parameters (bulk Lorentz factor,
$\Gamma$, electron's temperature, $\theta_e$ and magnetic field, $B$):
$E_{peak} \propto \Gamma \theta_e^2 B$. Given the large differences
among the various GRBs, it is difficult to explain the observed narrow
clustering of the peak energy without fine tuning the model's free
parameters.

These observational and theoretical drawbacks of the purely
non-thermal emission models have led to renewed interest in thermal
models \cite{EL00, MR00, DM02, MRRZ02, RM05, Ryde+06}. A key
difference between these new thermal models and earlier models is the
realization that a thermal component is not the sole emission
component, and is in most cases accompanied by a non-thermal
component. One difference between the different theoretical ideas is
the relative strength of the thermal vs. non-thermal parts of the
overall non-thermal spectra. Within the framework of the basic
``fireball'' model, such differences are explained by the different
photospheric radii: during the coasting phase, the photons suffer a
substantial adiabatic losses, and both the temperature and the thermal
luminosity drop as $\propto r^{-2/3}$. As the photospheric radius is
uncertain, weak thermal signal can be explained as originating from
large photospheric radius.

In many of these hybrid models, the observed sub-MeV peak was thought
to originate from the thermal component, while the non-thermal part
acts to broaden the ``Planck'' spectra. These assumptions enabled these
models to overcome many of the drawbacks of the pure thermal and pure
non-thermal models. In particular: (1) the existence of hard low
energy spectral slopes; (2) temporal variations in the spectral shape;
(3) the observed spectral correlations between the peak energy and
luminosity; (4) the high efficiency (thermal photons originate
directly from the explosion, and as such no kinetic or magnetic energy
dissipation is needed in producing them); and (5) they naturally
explain the existence of a complex spectral shape.

Despite these successes, these models were, by large, heuristic in
nature. For example, the thermal and non-thermal parts were treated as
a completely separate entities. Another example is that temporal
evolution was not treated in a quantitative way. In fact, a major
breakthrough took place when it was realized \cite{Ryde04, Ryde05}
that one needs to look at {\it time-resolved } spectra. As the
emission mechanisms vary with time, time-integrated spectra can easily
smear out any signal. Clearly, this produces a much larger technical
challenge. In many GRBs (certainly in the BATSE era), there were
simply not enough photons observed to enable spectral analysis on
short times. Still, there were several bright GRBs for which such an
analysis could be done.

\section{Observational status}
\label{sec:3}

A key point which is particularly confusing to many people, is the
fact that one needs to discriminate between direct and indirect
evidence for the existence of a thermal component. A direct evidence
for a thermal component - namely, a direct observation of black body,
or grey body spectra in GRBs is relatively rare. Furthermore, in most
cases in which it is observed, it is accompanied by a
non-thermal part. Nonetheless, the fraction of GRBs in which a pure
thermal component is observed increases with (1) the GRB brightness;
and (2) when time resolved spectroscopy is performed. These two facts
strongly points towards the possibility that in many bursts this
component is simply smeared, due to the low number of photons
detected: nearly by definition, the flux of most bursts observed is
close to the detector's limit, and thus only very few photons are
observed for most GRBs reported. Combined with the fact that both the
thermal and non-thermal parts of the spectra vary with time, it is
clear why direct observation of thermal component is difficult and
rare.

As opposed to this, indirect evidence for the existence of thermal
component exist in a very large fraction of bursts. These evidence are
based mainly on fitting the low energy spectral slopes (below the peak
energy) in ``Band'' model fits. The theory of synchrotron radiation
provides a robust upper limit on the low energy spectral slope that
can be observed: $F_\nu \propto \nu^{-\alpha}$, with $\alpha \geq 1/2$
(in the ``fast cooling'' regime) \cite{RL79, SNP96, SPN98}. This upper
limit, though, is {\bf lower} than the spectral slopes observed in the
majority of GRBs \cite{Crider+97, Preece+98, preece+02, GCG03}. This
result implies that the (simple version) of the synchrotron model by
itself cannot explain the spectra.  This motivated several authors to
suggest modifications to the model, by altering one or more of its
underlying assumptions. Suggestions include addition of a spatial
scale for the decay of this field \cite{PZ06, Zhao+14, UZ14, UZ15},
modify the acceleration process \cite{Murase+12b, AT15} or substantial
modification of the low energy particle distribution due to
inverse-Compton scattering \cite{Daigne+11}. When such modifications
are made, much better fits to the spectra can be made
\cite{Zhang_BB+16}.

Nonetheless, recent works suggested an even more robust way of testing the
synchrotron prediction, by looking at the spectral width of the entire
spectrum (rather than focusing on the low energy spectral slope).  The
results of these works \cite{AB15, Yu+15b}, in fact, suggest that a
thermal emission component exists in nearly 100\% of all GRBs.

\subsection{Direct evidence}
\label{sec:3.1}

The most robust way of claiming the detection of a thermal component
is by fitting the spectra - or part of the spectra, with a ``Planck''
(or modified Planck) function. This provides a direct probe of (1) the
thermal flux, and (2) the temperature of the thermal
component. However, unfortunately, as is stated above, this is often
not an easy task. 

There are two reasons for this difficulty, and both are related to the
analysis method.  First, for most GRBs the spectral analysis is based
on analyzing flux integrated over the entire duration of the prompt
emission, namely the spectra is time-integrated. Clearly, this is a
trade off, as enough photons need to be collected in order to analyze
the spectra.  For weak bursts this is the only thing one can
do. However, obviously this inevitably leads to smearing of any
time-dependent signal, and produce artifacts \cite{BR15}.

Second, the analysis is done by a forward folding method, which
implies the following: a model spectrum is convolved with the detector
response, and compared to the detected counts spectrum. The model
parameters are varied in search for the minimal difference between
model and data; this gives the best fitted model's
parameters. However, this method implies that the outcome of the
analysis are biased by the initial hypothesis: as the fitted results
depend on the model that was initially chosen, {\bf two different
  models can provide equally good fits to the same data}.

A particular source of difficulty is the wide-spread use of the
``Band'' model in fitting the prompt emission spectra. Having only 4
free parameters, fitting the data using this model is unable to
capture any ``wiggles'' or ``bumps'' that may exist in the
spectra. But the existence of such wiggles are exactly the indicators
for a possible thermal component atop a non-thermal spectrum! In
particular, a composite spectra, in which a thermal component is only
one ingredient would be completely smeared. This is the key reason why
study of a thermal component in GRBs was delayed for over a decade.

\begin{figure}[t]
\begin{center}
\includegraphics[width=2in]{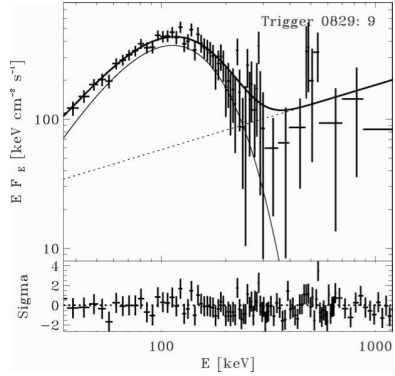}
\end{center}
\caption{A ``hybrid'' model fit to the spectra of GRB 910927 detected by BATSE.}
\label{fig:1}
\end{figure}

In order to overcome both problems, a different analysis method was
suggested by F. Ryde \cite{Ryde04, Ryde05}. Firstly, fits to the
prompt emission spectra were carried using a ``hybrid'' model, that
contains a thermal component (a ``Planck'' function) in addition to a
single power law (see Figure \ref{fig:1}). The rational was to keep the number of free
parameters to four (same as in the ``Band'' model), for ease of
comparison of quality of fits. On the downside, clearly a single power
law cannot represent any physical emission process; it can, though, be
acceptable over a limited energy range, as was available during the
BATSE era. A second novelty was the use of time resolved spectral
analysis. While this limited the number of bursts in which the
analysis could be carried to only $\mathcal{O}(10)$, it enabled, for
the first time, the detection of a temporal evolution of the
properties of the thermal component (see Figure \ref{fig:2}).

\def\figsubcap#1{\par\noindent\centering\footnotesize(#1)}
\begin{figure}[b]%
\begin{center}
  \parbox{2.1in}{\includegraphics[width=2in]{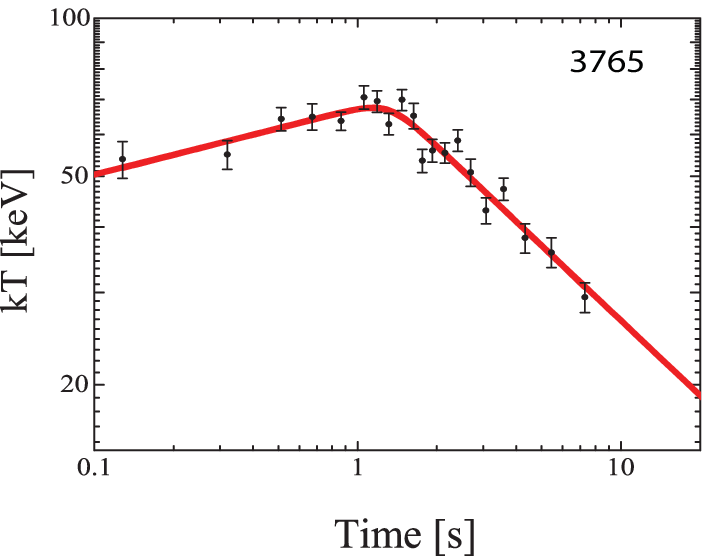}\figsubcap{a}}
  \hspace*{4pt}
\parbox{2.1in}{\includegraphics[width=2.4in]{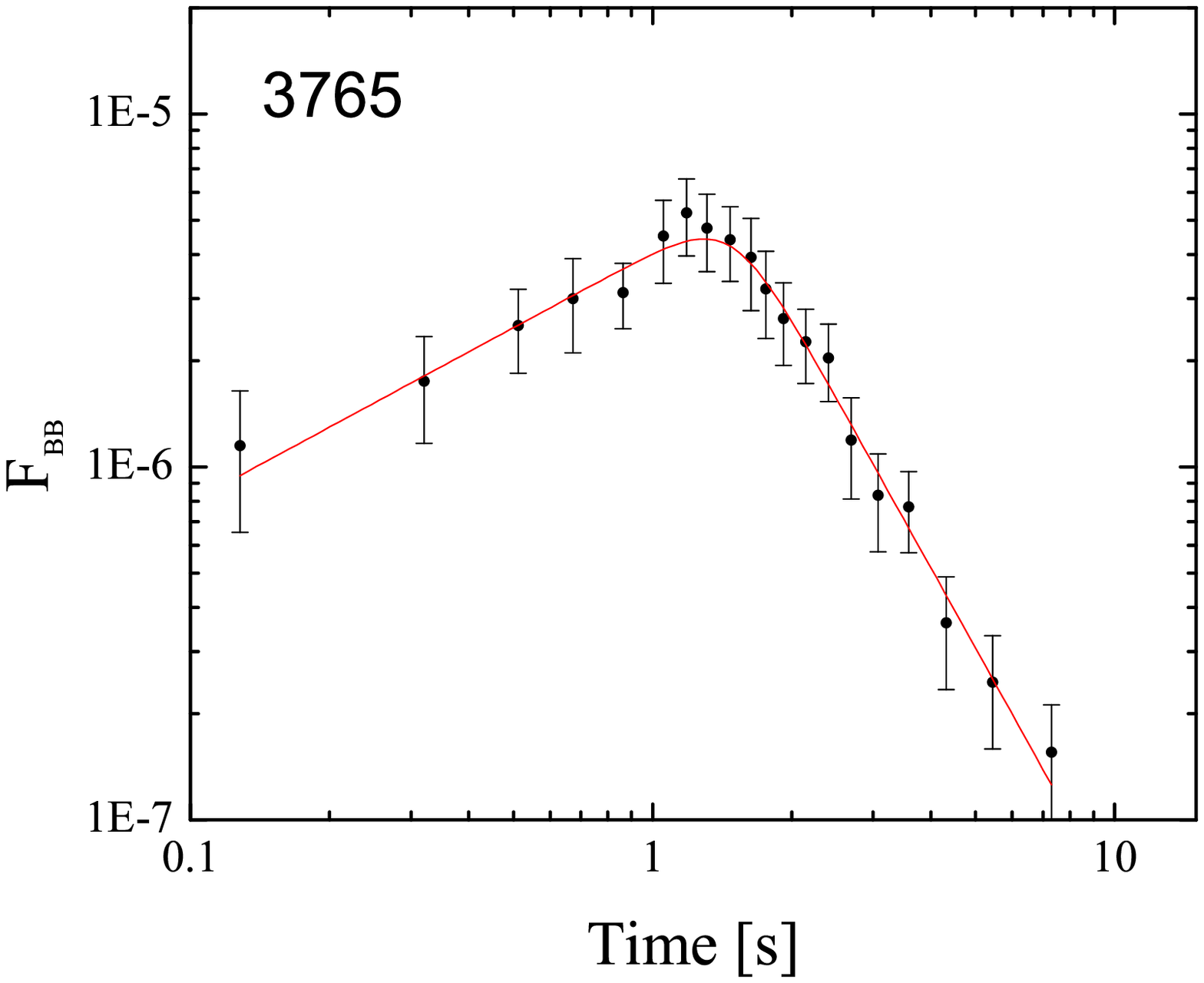}\figsubcap{b}}
  \caption{Temporal evolution of the temperature (left) and flux (right) of the thermal component of GRB950818.
The temperature is nearly constant for $\sim 1.5$~s, afterwards it decreases as a power law in
time. The flux also shows a broken power law temporal behavior, with a break time which is within the errors of the break time in
the temperature evolution. 
}
  \label{fig:2}
\end{center}
\end{figure}

\begin{figure}[t]
\begin{center}
\includegraphics[width=8cm]{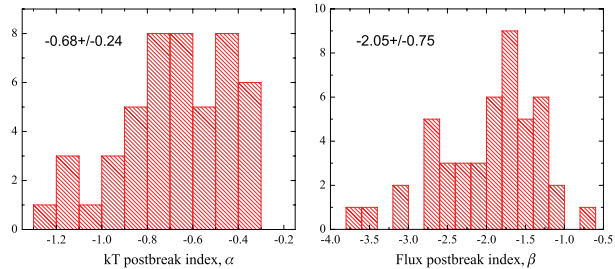}
\end{center}
\caption{Histogram of the late time (after the break) temporal
  evolution of the temperature (left) and flux (right) in the sample
  of 56 GRBs from Ref. \cite{RP09}.}
\label{fig:3}
\end{figure}

This analysis was extended by Ref. \refcite{RP09}, to study the
properties of 56 BATSE GRBs, the largest sample at that date. That
analysis revealed a clear repetitive behavior in the properties of
the thermal component, which were found to be distinctive from those
of the non-thermal part. The temporal evolution of both the
temperature and thermal flux show a well-defined, broken power law
behavior. The temperature was found to be nearly constant at $t \leq
t_{brk} \sim$few~s, and then decayed as $T(t) \propto t^{-\alpha}$,
with $\langle \alpha \rangle = 0.68$. The thermal flux first rise as
$F(t) \propto t^\beta$ with $\langle \beta \rangle = 0.63$, and after
break time (which typically coincide with the break time observed in the
evolution of the temperature) it decays with an average index $\langle
\beta \rangle = -2$ (Figure \ref{fig:3}). This repetitive temporal
behavior serves as an independent indicator for the existence of a
distinct thermal component; for the least, a component that is
distinct from the non-thermal part of the spectrum in both its
spectral and temporal properties (see discussion on the theoretical
interpretation below).

The launch of {\it Fermi gamma-ray space telescope} in 2008, enabled
much broader spectral range than was available prior to its launch.
The increase in spectral range made it clear that in many GRBs a
simple 4 component ``Band'' function is insufficient to describe the
broad band spectrum. Contrary to some initial expectations
\cite{Granot+10}, it was found that in many GRBs the prompt spectra
can only be fitted with multiple spectral components, that contain
several ``wiggles''. In some basic sense, this is of no surprise, as
the ``fireball'' model was constructed to enable multiple emission
zones with different physical conditions. There is thus no a-priori
reason to believe that all emission zones will produce identical
spectra. This conclusion is further strengthen by {\it Fermi's}
detection of a typical delay of few seconds in observations of the
high energy photons [e.g., \cite{Abdo+09a, Ackermann+10, Ackermann+13,
    Ackermann+14}].

Despite this progress, still in the {\it Fermi} era it is found that
the ``Band'' function provides reasonable fits to the vast majority of
GRBs \cite{Zhang+11, Goldstein+12, Bosnjak+14, Gruber+14,
  VonKienlin+14, Yu+16}. Many of these fits differ than earlier fits
by the use of time-resolved analysis.  Still, as explained above, in
these fits ``Band'' model template was a-priori assumed, implying that
``wiggles'' could not be detected. Thus, these fits cannot exclude the
possibility that thermal component does exist, and could be revealed
if more sophisticated templates were in use. Furthermore, it is found
that the deviations from the ``Band'' fits are more likely to occur in
bright GRBs \cite{Tierney+13}, emphasizing the important of sufficient
photon number statistics in drawing conclusions about GRB spectral
properties.  Nonetheless, these analyses imply that in most GRBs the
observed spectrum is, by large, non-thermal and the thermal component,
if indeed exist, is not dominant - it is always accompanied by a
non-thermal part.

\begin{figure}[t]
\begin{center}
 \includegraphics[scale=0.3,bb=-40 10 550 690,clip]{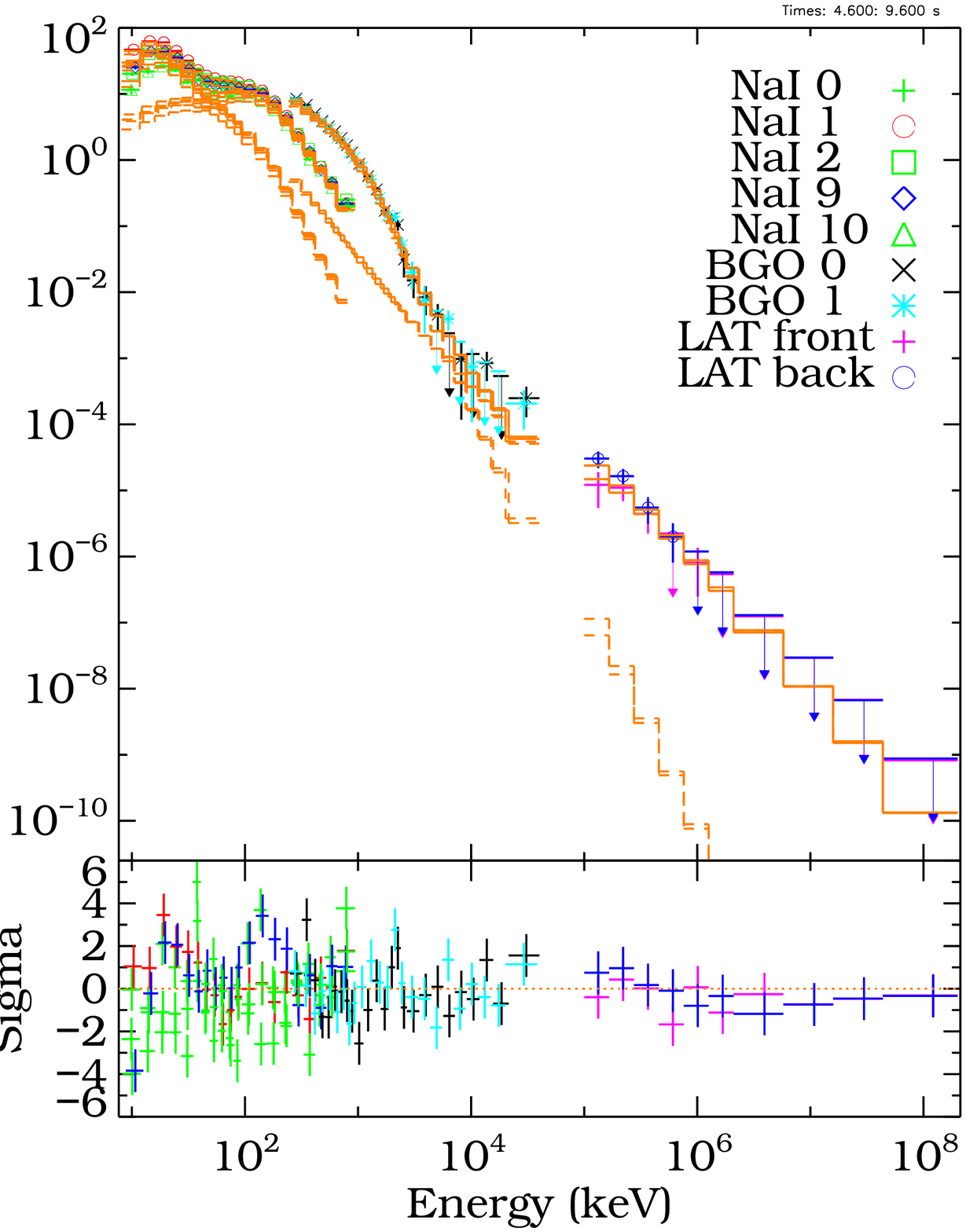}
  \includegraphics[scale=0.3,clip]{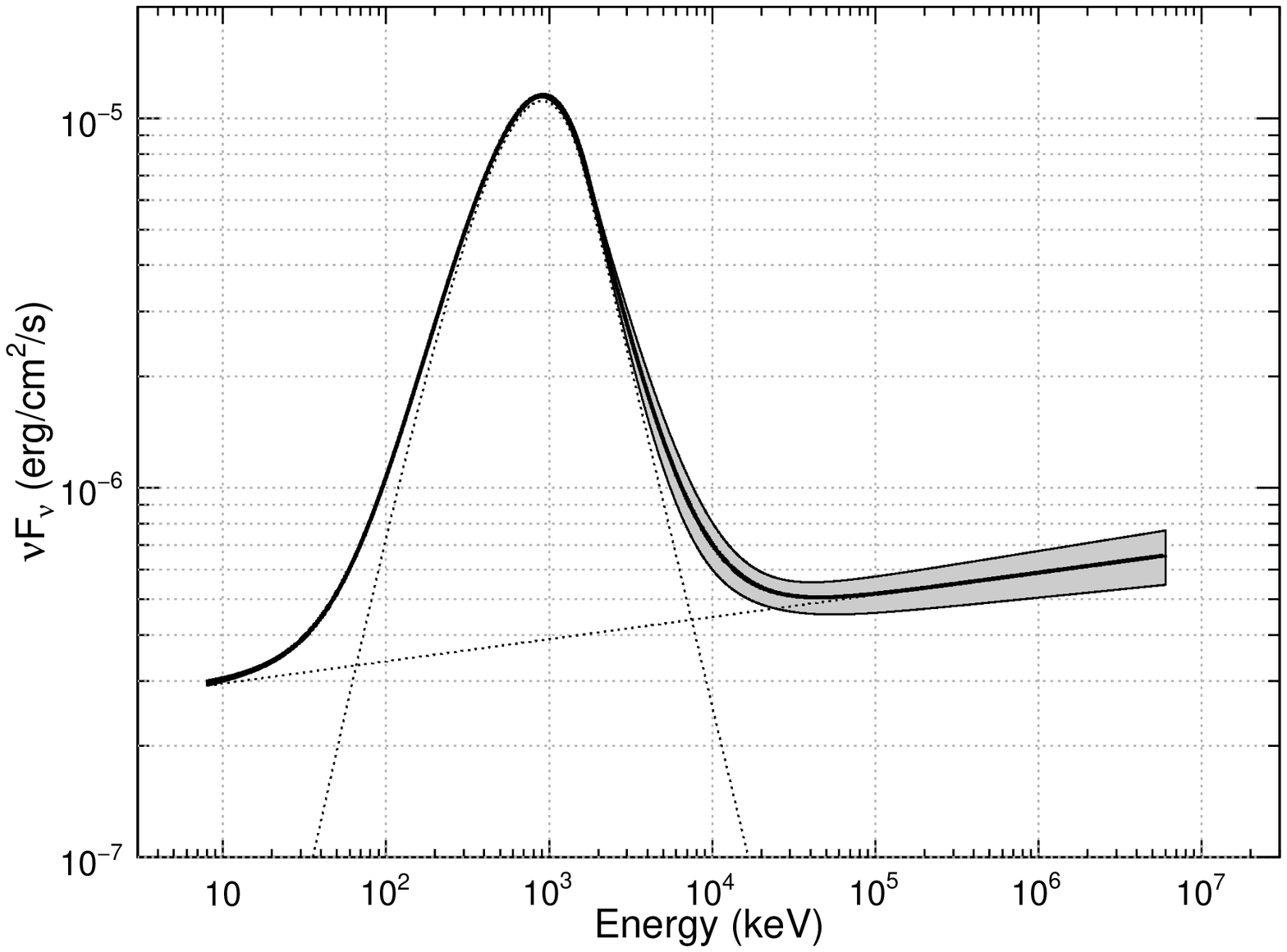}
\end{center}
\caption{Spectrum of GRB090902B at 4.6 - 9.6 seconds after the trigger show clear thermal component. Left: countes spectrum. Right: unfolded  $\nu F_\nu$ spectrum. Figure adopted from Ref.\cite{Abdo+09a}. }
\label{fig:4}
\end{figure}

An exceptional burst was the very bright burst GRB090902B
\cite{Abdo+09a, Ryde+10, Ryde+11}. Its spectra showed an extremely
bright thermal component, well distinct from the non-thermal part
(see Figure \ref{fig:4}). The thermal component was so pronounced,
that the spectra rejected any attempt to be fitted with a ``Band''
model, despite numerous efforts. Attempts to fit only the thermal part
with a ``Band'' model concluded that both the low and high energy
slopes are so steep that only a ``Planck'' spectrum could provide a
physical origin to the fit. Furthermore, being so bright, it was easy
to follow the temporal evolution of the prompt emission. After a few
seconds, the thermal component began to spread, resembling more and
more a ``standard'' ``Band'' function \cite{Ryde+11}.

While GRB090902B was a single event, its appearance clearly
demonstrated the fact that multiple spectral components indeed exist
in the prompt emission spectra. Furthermore, their relative strength can vary
during the prompt phase. These findings encouraged several authors to abandon the
``Band'' fits and search for thermal emission, by modifying the
a-priori assumed template.  Indeed, once more complicated templates
began to be used, a thermal component (on top of a non-thermal
spectra) was found in several bursts. A few notable ones are GRB090510
[\cite{Ackermann+10}], GRB090618[\cite{Page+11, Izzo+12}], GRB110721A
[\cite{Axelsson+12, Iyyani+13}], GRB100724B [\cite{Guiriec+11}],
GRB100507 [\cite{Ghirlanda+13}], GRB120323A [\cite{Guiriec+13}],
GRB110920A [\cite{Iyyani+15}], and GRB101219B [\cite{Larsson+15}].

\begin{figure}
\begin{center}
\includegraphics[width=8cm]{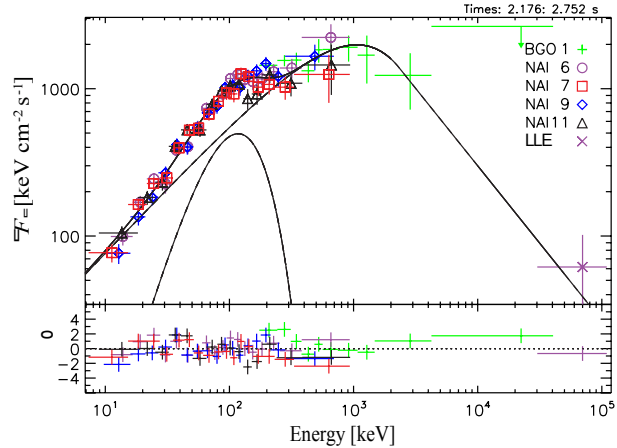}
\end{center}
\caption{The spectra of GRB110721A is best fit with a ``Band'' model (peaking
at $E_{peak} \sim 1$~MeV), and a blackbody component (having
temperature $T \sim 100$~keV). The advantage over using just a
``Band'' function is evident when looking at the residuals (Taken from
\cite{Iyyani+13}) }
\label{fig:5}
\end{figure}

Many of these fits used an ``advanced'' version of Ryde's original
``hybrid'' model, by modeling the non-thermal part of the spectra with
a ``Band'' function. This implies that these models are semi-physical
(the thermal part has clear physical origin, while the origin of the
``Band'' function is unclear). A repeated result is that, using these
fits, the thermal component often {\bf does not} coincide with the
peak energy, but is observed as a lower energy ``wiggle'' on top of
the ``Band'' low energy spectral slope (see Figure \ref{fig:5}). The
observed thermal flux is at the range of tens of \% of the total flux.
This result, though may be attributed to a selection bias: if the
fraction of thermal photons was lower than that, they could not have
been detected. Recetly, it was shown that the ``Band'' function in
these fits can be associated with (slow cooled) synchrotron emission,
at least in a few bursts \cite{Burgess+14, BRY15}

There are therefore two key conclusions from these works. 
\begin{enumerate}
\item The main conclusion is that thermal component is most likely
  ubiquitous. A main reason why it is not observed in many bursts is
  simply that many fits are done using templates that don't enable its
  discovery. A secondary reason is that (by definition) most bursts
  will be seen with very low photon statistics, that will not enable
  its detection.
\item In most cases in which a thermal component was detected, it was
  accompanied by a non-thermal part, which typically was seen to be
  with higher flux. Furthermore, for most bursts in which a thermal
  component is observed, the peak of the thermal component does not
  coincide with the spectral peak. In several cases, the observed peak
  of the ``Band'' spectrum is too energetic to be explain by a thermal
  component \cite{Zhang+12}.
\end{enumerate}

\subsection{Thermal emission observed at late times}

A few authors have reported a thermal component that was detected not in
the $\gamma$-rays, but rather in the X-ray band. In these bursts, the
typical temperature of the thermal component is at the $\sim$~keV
range, as opposed to $\sim 100$~keV observed in the bursts discussed
above. Furthermore, in these bursts the thermal component was observed
to last hundreds of seconds, extending well into the afterglow
phase. In this category, there are both low luminosity GRBs such as
GRB060218 [\cite{Campana+06, Shcherbakov+13}] or GRB100316D
[\cite{Starling+11}] but also many GRBs with typical luminosities,
e.g., \cite{Page+11, Starling+12, SpS12, FW13, Bellm+14, Schulze+14,
  Piro+14, BaRa15}.

Common to all GRBs in this category are (1) the fact that the thermal
component is observed well into the afterglow phase; and (2) the
inferred values of the Lorentz factors are at least an order of
magnitude lower than that of ``standard'' GRBs: $\Gamma \lesssim$ few
tens, and in some cases much lower, $\Gamma \gtrsim 1$. These results
indicate that the physical origin of the thermal component in these
bursts may be different than those bursts in which a thermal component
is seen at higher energies. Leading models are supernovae shock
breakout and emission from the emerging cocoon (which we will briefly
discuss below).  Overall, these detections imply that there is more
than a single way of producing a thermal component in GRBs.

\subsection{Indirect evidence}
\label{sec:3.3}

As explained above, direct detection of a thermal emission
(observation of a ``Planck'' function) is still relatively rare. A
major part of this is the use of the ``Band'' fitting function. A
second reason is the lack of sufficient photon statistics; and a third
possible reason (which will be described below) is possible various
physical mechanisms that act to smear the pure ``Planck'' function. In
fact, such mechanisms, if indeed operate, will make it very difficult
to claim any detection of the thermal component.

Nonetheless, there are ample of indirect evidence for the existence of
a thermal emission component in the vast majority of bursts. These
evidence are based on the fact that the ``Band'' fit, by itself, does
not contain any physical meaning. Thus, it needs to be interpreted in
the framework of one (or more) known radiative mechanisms. Clearly,
there is more than one way of interpreting the data. Nonetheless, some
of the key observed properties cannot be explained in the framework of
any of the alternative radiative mechanisms, or that they require
``fine tuning'' of the model parameters. These same properties can, in
some cases, be much easier explained if one assumes that the observed
spectrum is composed of a (modified) Planck component.

Among the key observed properties, one finds the following: 

\begin{enumerate}
\item{\bf Steep low energy spectral slopes}. As discussed above, when
  fitting the prompt spectrum with a ``Band'' function, the average
  low energy spectral slope obtained is $\langle \alpha \rangle = -1$.
  The low energy spectral slope of about 85\% of the GRBs fitted in
  this way are found to be steeper than $-1.5$, which is the upper
  limit allowed by synchrotron emission (in the ``fast cooling''
  regime) \cite{Crider+97, Preece+98, preece+02, GCG03, Kaneko+06,
    Nava+11a, Goldstein+12, Goldstein+13, BRY15}. This result therefore
  implies that the (optically thin) synchrotron emission cannot be
  responsible to this emission, at least without significant
  modifications to its underlying assumptions.
\item {\bf Spectral width.} Recent works measured the spectral width
  of the prompt emission. The advantage of this method is that it is
  not sensitive to any assumptions about the initial particle
  distribution (power law or not) or to the question of fast vs. slow
  cooling. The results of these works [Ref. \cite{AB15, Yu+15b}]
  clearly indicate that synchrotron emission - even from a Maxwellian
  distribution of particles produces a much wider spectrum than is
  observed. We point out that a full assessment of these results
  require direct fits to the synchrotron model, in order to overcome
  the ambiguity caused by the forward-folding technique.
\item {\bf Observed correlations.} Several correlations have been reported
  between the peak energy and the total energy or luminosity of a GRB
  [e.g., \cite{Golenetskii+83, Amati+02, Ghirlanda+04,
      Yonetoku+04}]. Within the framework of the ``synchrotron''
  model, the peak energy has a strong dependence on the free model
  parameters, $E_{\rm peak} \propto \Gamma \theta_{el}^2 B$, where
  $\Gamma$ is the bulk Lorentz factor, $B$ is the magnetic field and
  $\theta_{el}$ is the characteristic electron's temperature. Thus,
  first, there is no a-priori reason why the peak energy in all GRBs
  should be roughly at the sub-MeV range; the values of all these
  three parameters can vary substantially between the
  bursts. Furthermore, there is no a-priori simple explanation to the
  observed peaks, without adding further assumptions \cite{ZM02,
    LZ04}.

Explaining the observed peak with a thermal emission also requires
tuning of the parameters [e.g., \cite{MR00, RM05, Peer+12}], though it
seem to be less restrictive than in the synchrotron model. Similarly,
while thermal emission by itself cannot explain any of the observed
correlations, it can do so with a relatively minor modifications
[e.g., \cite{TMR07, Fan+12, Lazzati+13, Guiriec+15}].
\end{enumerate}

\section{Theory}
\label{sec:4}

\subsection{Existence of a thermal component}
\label{sec:4.1}

As discussed in section \ref{sec:2} above, a thermal emission
component was expected from the very early days following the
realization that GRBs are cosmological.  The huge optical depth at
the base of the flow ($\tau \sim 10^{15}$) implies that any existing
radiation will thermalize before escaping. Thus, a thermal component
is an inherent part of the cosmological ``fireball'' model that was
expected very early on \cite{Goodman86, Pac86,Pac90}. This conclusion
is not changed if the acceleration is mediated not by photons but by
reconnection of strong magnetic fields that may exist in the innermost
regions of the outflow \cite{Thompson94, SDD01, Drenkhahn02,
  DS02}. None of the early models, though provided any robust
predictions about the relative importance of the thermal vs. the
non-thermal part. The expected thermal to non-thermal flux ratio
depends on several unknown factors such as the radius of the
photosphere, the radii of the energy dissipation episodes that lead to
the emission of the non-thermal radiation, and the efficiency in
producing the non-thermal photons. As a result, a variety of models
exist, in which the relative strength of the thermal component vary -
from dominant \cite{EL00, DM02} to sub-dominant \cite{MR00, MRRZ02,
  RM05}.

Lacking any clear theoretical prediction, the values of the unknown
models parameters were deduced from observations. The lack of a clear
thermal signal in the observed data therefore led many people to
consider a parameter space region in which a thermal component is
sub-dominant. This can be easily obtained if the photospheric radius,
$r_{ph}$ is large enough, so that the thermal photons suffer
substantial adiabatic losses prior to their escape. In such a
scenario, it was concluded by many authors that the origin of the
observed peak energy at the sub-MeV range is most likely due to
synchrotron emission [for a very partial list, see Refs. \cite{RM92,
    MR93, MLR93, MRP94, PX94, PM96, Tavani96, Cohen+97, SP97b, PL98,
    DM98}]. This line of reasoning was forced to be reconsidered once
evidence began to accumulate for the inconsistency of the observed
spectra with the theoretical predictions of the synchrotron model, in
the early 2000's.

Even in those versions of the GRB ``fireball'' model in which only a
small number of thermal photons is initially assumed, the theory
allows the creation of a large number of thermal photons at a later
stage. These photons can be created by friction between the jet
components, or the jet component and the surrounding material. For
example, free neutrons are expected to decouple the protons at small
radii, below the photosphere, due to the lower cross section for
proton-neutron collision relative to Thomson cross section. The
friction between the neutrons and protons can result in energy
dissipation, that eventually heats the jet and is capable of producing
a large number of thermal photons \cite{Beloborodov10, Vurm+11}
(provided, of course that this energy dissipation occurs sufficiently
below the photosphere). It was further proposed that these photons may
be associated with the peak energy \cite{Beloborodov13}.

The association of long GRBs with core-collapse supernovae of type
Ib/c \cite{Galama+98, Hjorth+03, Stanek+03, Campana+06, Pian+06,
  Cobb+10, Starling+11} led to the conclusion that at least long GRBs
are associated with the death of massive star. In this so-called
``collapsar'' model \cite{Woosley93, Pac98, Pac98b, Fryer+99, MW99,
  Popham+99, WB06}, the GRB jet drills its way through the collapsing
material \cite{Aloy+00, MacFadyen+01}. Interaction between the jet and
the stellar envelope will lead to the formation of shock waves which
will heat the plasma. These shock waves could potentially occur below
the photosphere, thereby providing another channel for producing
thermal photons \cite{Lazzati+09, Morsony+10}. A prediction of this
model is the association of the thermal (and non-thermal) emission
time with the time it takes the stellar material to collapse, which is
of the order of $\sim 10$~s \cite{Aloy+00, Morsony+07, MA09,
  Bromberg+11a}. Once the jet completes its crossing through the
stellar envelope, the external pressure rapidly drops, and the
radii of these recollimation shocks would gradually increase until
they would eventually disappear and the flow becomes free. However,
this stage typically lasts a duration of a few sound crossing
times, $\simeq 10$~s \cite{Morsony+07, MA09, Lopez+13}. During this
epoch, these shocks are roughly at their initial location, thereby are
capable of producing a thermal emission component.

On its way out of the collapsing star, the jet heats and pushes the
collapsing stellar material, forming a hot ``cocoon'', that expands
outside of the stellar envelope following the emergence of the
jet. This hot cocoon, which is much slower than the jet itself
(estimates based on numerical models reveal $\Gamma_c \sim 10$) is
optically thick, with optical depth that can reach few hundreds
\cite{PMR06b}. It may therefore be responsible for the late time
thermal emission observed \cite{Toma+09}. Finally, additional source
of thermal emission may be the interaction of the relativistic GRB jet
with the supernova shell \cite{Thompson06}. 

Thus, to summarize this section, in fact there is a consensus that a
thermal emission component should exist in cosmological GRBs; this is
agreed by many different models that consider different dynamical
scenarios. None of currently existing models, though, give any
robust prediction on the expected strength of the thermal component,
and the models differ in the relative importance of this
component. Lacking a clear theoretical prediction, in fact in nearly
all models the role of a thermal component is left as a free parameter
that is scaled by observations.

\subsection{Broadening of the thermal components}
\label{sec:4.2}

The fact that thermal emission was predicted to exist (and in some
models predicted to be dominant) in the GRB prompt spectra, naturally
raise the question of its lack or weakness in the observed
spectra. One immediate answer for its lack is adiabatic energy losses
below the photosphere. As discussed above, these are expected in
parameter space region in which the photospheric radius is very
large. In such a case, the thermal photons loose their energy below
the photosphere at the expense of the plasma's bulk kinetic
energy. As a result, when the thermal photons decouple the plasma at
the photosphere, both their temperature and thermal flux are low -
close to, or even below the detection limit. This view was the leading
view up until the first half of the 2000's, and is still a leading
view by several scientists.

In this scenario the dominant emission processes responsible for the
observed signal therefore take place way above the photosphere. As
such, they must be non-thermal in nature: the leading mechanisms are
synchrotron, inverse Compton or, alternatively, emission from
energetic hadrons.  These processes follow episode(s) of energy
dissipation (either kinetic or magnetic), which accelerate particles
that produce the non-thermal radiation. According to this picture, the
thermal component plays a very small or negligible role in shaping the
observed spectra.

As explained above, this view was challenged in the early 2000's by
various observations that were found to be in contradiction to the
optically thin emission model predictions. One branch of solutions
was, and still is, to modify one or more of the underlying assumptions
of the optically thin models (see discussion in section \ref{sec:3}
above). An alternative approach is to look at mechanisms that
may modify the thermal component itself in such a way that the modified
spectral shape will resemble the observed one. If this line of
reasoning is correct, the thermal emission component in fact plays a very central
role in determining the observed spectra. The observed GRB spectra
deviates from a ``Planck'' function (and thus seem as being non-thermal) due to various
physical processes and geometrical effects. 

In this section we discuss some possible mechanisms that can act to
modify the Planck spectra and their implications. Of course, if this
is the correct scenario, it is much more difficult to prove the
existence of an initial thermal component from the observed signal.

\subsubsection{Physical broadening and connection with the non-thermal spectra}
\label{sec:4.2.1}

The most natural way of modifying a ``pure'' thermal component is by
assuming that some part of the available energy (kinetic or magnetic)
is dissipated below, or close to the photosphere. In fact, this is a
natural part of the classical ``fireball'' model, in which the jet's
kinetic energy is dissipated by instabilities in the outflow that lead
to internal shock waves \cite{RM94, Meszaros06}. As we pointed out
above, the ``fireball'' model (in all its different versions) does not
provide strong constraints on the radii of the internal collisions
between the outflow components that dissipate its kinetic energy. Part
of these collisions may very well occur below the
photosphere. Similarly, in models in which the outflow is highly
magnetized, it is often assumed that the magnetic energy is dissipated
at a constant rate from the fast magnetosonic radius onward
\cite{SDD01, LK01, Drenkhahn02, DS02}, implying that part of the
energy is dissipated below (but close to) the photosphere.

This dissipated energy is used (at least in part) to heat and/or
accelerate plasma particle (electrons and possibly protons). A leading
mechanism by which this dissipation can occur is by sub-photospheric
(radiation-mediated) shock waves. The microphysics of particle
acceleration in shock waves is of yet an open question.  It was
recently argued that sub-photospheric shock waves lack the the
structure that enable the acceleration of particles to high energies
\cite{Levinson12}. This, however, is expected to have only little
effect on the emerging spectra, with respect to a scenario in which
the particles are thermally heated by the shock waves.  The reason is
as follows. Once the particles are heated or accelerated, they
radiatively cool extremely rapidly by upscattering the thermal
photons. Due to the fact that below the photosphere the number of
thermal photons in the plasma is much greater than the number of
particles, any energetic particle undergoes very many scattering, and
therefore its cooling time is many orders of magnitude shorter than
the dynamical time \cite{PMR05}. This means that the energetic
particles will form a (quasi-) steady state very rapidly, which could
be characterized by a (quasi-) Maxwellian distribution. Their
temperature is determined by balance between the heating (whose
details depend on the unknown details of the dissipation mechanism) on
the one hand, and radiative cooling on the other hand. As long as the
external heating is active, the particle's temperature will inevitably
be higher than the temperature of the thermal photons in the plasma,
that are not directly affected by the heating process.  The result is
the formation of a 'two temperature' plasma, containing a population
of thermal photons with (comoving) temperature $T'_\gamma$, and a
population of hotter electrons, characterized by a higher temperature,
$T'_{el} > T'_\gamma$.

As was discussed in Ref. \cite{PMR05}, the particle's temperature is
highly regulated, and is very weakly depending on the model's
parameters.  It depends on only two parameters: (i) the ratio between
heating rate and cooling rate (or, alternatively the energy density in
the particles and the thermal photons), and (ii) the optical depth in
which the dissipation takes place, which is governed by the radius of
energy dissipation. The optical depth determines the number of
scattering. For optical depth at the range $1 \lesssim \tau \lesssim
100$, the electron's steady state (comoving) normalized temperature is
$k T'_{el} / m_e c^2 \sim 0.1 - 1$.

\begin{figure}
\begin{center}
\includegraphics[width=10cm]{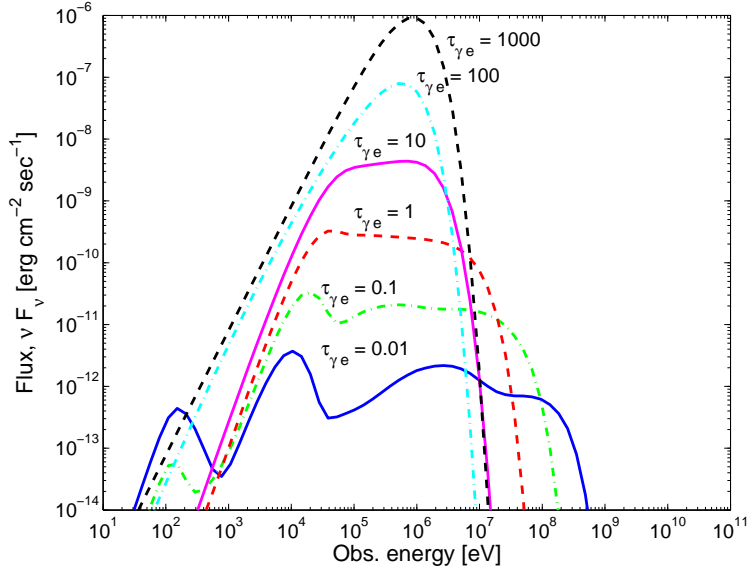}
\end{center}
\caption{Time averaged broad band spectra expected following kinetic
  energy dissipation at various optical depths. For low optical depth,
  the two low energy bumps are due to synchrotron emission and the
  original thermal component, and the high energy bumps are due to
  inverse Compton phenomenon.  At high optical depth, $\tau
  \geq 100$, a Wien peak is formed at $\sim 10 \keV$, and is
  blue-shifted to the MeV range by the bulk Lorentz factor $\simeq
  100$ expected in GRBs.  In the intermediate regime, 
$0.1 < \tau < 100$, 
a flat energy spectrum above the thermal
  peak is obtained by multiple Compton scattering. Figure taken from Ref. \cite{PMR06}}
\label{fig:6}
\end{figure}

The electron's distribution settle to the quasi steady state on a time
scale much shorter then the dynamical time. Thus, during most of the
dynamical time, the hotter electrons up-scatter the thermal photons,
forming a secondary distribution at energies above $T'_\gamma$
\cite{PMR06, LB10, Toma+11, VM12, Hascoet+13, CL15}. This is
demonstrated in Figure \ref{fig:6}, taken from Ref. \cite{PMR06}. 

The resulting spectral shape depends on the optical depth in which the
dissipation takes place. This is most easily understood when looking
at the two extremes. If the radius at which the dissipation occurs is
much greater than the photosphere ($\tau \ll 1$), then the thermal
photons will have very few interactions with the energetic particles,
and will be observed as an independent spectral component. Some
thermal photons would serve as seed photons for Compton scattering by
the energetic particles; the relative strength in the observed spectra
would depend on the Compton $Y$ parameter. In such a scenario, the
energetic particles (that may be energized already below the
photosphere; see \cite{Ioka+07}) will radiate non-thermal
emission. Two additional peaks may therefore be seen - due to
synchrotron emission at lower energies, and synchrotron self Compton
(SSC) peak at higher energies. 

At the other extreme, in which the dissipation occurs in a very small
radius ($\tau \gg 1$), the up-scattered photons will have ample of
time to re-distribute their energy, and a new thermal distribution
would emerge; simply, the energy given to the particles by the
dissipation mechanism would be distributed among the particles and
photons. The resulting spectrum will be thermal. Interestingly, in
order for this to happen, it is enough that the dissipation takes
place in region in which the optical depth is greater than few
hundreds (see Figure \ref{fig:6}).

The most interesting signal is observed if the dissipation occurs at
intermediate values of the optical depth, $\tau \sim$~few - few
tens. The addition of hot particles below the photosphere implies that
some of the thermal photons will be up-scattered; but since by
assumption $\tau$ is not very high, full re-thermalization could not
be achieved. As explained above, since the number of photons is much
greater than the number of particles, each particle will undergo very
many scattering, and so the particle's distribution will be quasi
Maxwellian, as opposed to the photon distribution.

The initial thermal component is expected to somewhat weaken, as
thermal photons are up-scattered; though the thermal component
will maintain its original temperature. The main radiative process
above the thermal peak will be inverse-Compton scattering, by the
quasi thermal electrons. It is not hard to show that at the range
$T_\gamma < E < T_{el}$, the emerging spectra is a power law in
energy. For a relatively large parameter space region, the resulting
spectra will be flat \cite{PMR06, Giannios06, Giannios08}.  Additional
radiative mechanisms, such as synchrotron emission, may contribute to
the lower energy part of the spectrum (below the thermal peak). Thus,
in this case, one does not expect a continuation of the power law from
above the thermal peak to below it. This is consistent with the
negative results found when a search for a single power law extending
both above and below the thermal peak were conducted
\cite{Ghirlanda+07, Frontera+13}.

The results presented in Figure \ref{fig:6} are calculated for a
single dissipation episode.  In explaining the complex GRB lightcurve,
multiple such episodes (e.g., internal collisions) are expected. Thus,
in reality, a variety of observed spectra, which are superposition of
the different spectra that are obtained by dissipation at different
optical depth are expected \cite{KL14}.

\begin{figure}
\begin{center}
\includegraphics[width=8cm]{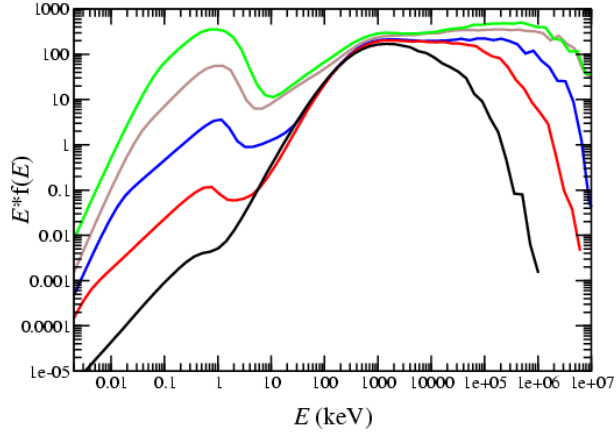}
\end{center}
\caption{Resulting spectra for dissipation occurring in highly
  magnetized models, as a function of the baryon load (or the
  magnetization).  From bottom to top the curves correspond to
  magnetization $\sigma_0=$ 40, 50, 60, 70, 100 (or corresponding
  baryon loading $\eta \simeq $ 250, 350, 460, 590, 1000)
  respectively. The high $\sigma _0$ flows are characterized broader
  spectra. The model predicts that bright prompt optical and UV
  emission is accompanied by powerful $\sim $GeV emission. For bright
  optical emission, the optical spectrum is expected to be
  hard. Figure taken from Ref. \cite{Giannios08}.}
\label{fig:7}
\end{figure}

The key results of this model do not change if one considers highly
magnetized plasma \cite{Giannios06, Giannios08, Giannios12, VM12,
    BP14}. A main difference between the highly magnetized models and
the radiative dominated ones is the assumption that the source of
energy that is used in heating the plasma is reconnection of magnetic
field lines. As opposed to internal shock waves which are discrete in
nature, the magnetic energy dissipation is expected to occur in a more
gradual way along the flow. Thus, in this model, gradual heating of
the plasma particles is expected from below the photosphere to above
it. The resulting spectra is surprisingly similar to the one obtained
in the discrete dissipation case; see Figure \ref{fig:7}, taken from Ref.
\cite{Giannios08}.

The model of sub-photospheric energy dissipation thus have four very
important advantages. First, it enables to explain some of the key
properties of the observed spectra that cannot be explained in the
framework of the optically thin, non-thermal emission models
\cite{Veres+12, Lopez+14}. Equally important is the fact that the
predicted spectra of this model are only weakly sensitive to many of
the uncertainties, such as the unknown outflow magnetization,
etc. This was investigated numerically by several authors for
different dynamical models \cite{Nagakura+10, Mizuta+11, MN12,
  Lazzati+13}, as well as magnetization parameter \cite{GZ15}. Third,
by slight modification of a single parameter - the optical depth in
which the dissipation (or most of it) takes place, the emerging
spectra can have very different shapes (see Figures \ref{fig:8},
\ref{fig:9}, taken from \cite{Ahlgren+15}). The sub-photospheric
dissipation model therefore unifies different spectra that seem to be
qualitatively different when fitted with a ``Band'' model into one
framework. It can further be tested by comparing high energies
spectral cutoffs \cite{GZ08}. Finally, as the origin of most of the
photons is thermal, the efficiency problem in kinetic or magnetic
energy conversion discussed above is much less severe. Most of the
radiated energy is already in the form of thermal photons, and the
dissipated energy acts to re-distribute them. Due to these advantages,
this model attracted a lot of attention in recent years [e.g.,
  \cite{Ioka+07, TMR07, Lazzati+09, LB10, Beloborodov10, Mizuta+11,
    LMB11, Toma+11, Bromberg+11, Levinson12, Veres+12, VLP13,
    Beloborodov13, Hascoet+13, Lazzati+13, AM13, DZ14, C-M+15,
    Santana+16}].
  
\def\figsubcap#1{\par\noindent\centering\footnotesize(#1)}
\begin{figure}[b]
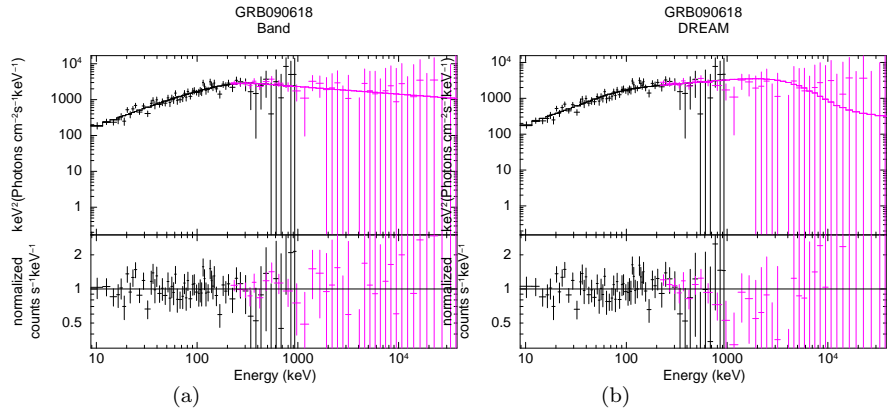
%
\begin{center}
 \parbox{2.1in}{\includegraphics[width=2in,angle=-90]{fig8a.eps}\figsubcap{a}}
  \hspace*{4pt}
\parbox{2.1in}{\includegraphics[width=2in,angle=-90]{fig8b.eps}\figsubcap{b}}
  \caption{Fits to GRB090618 at time bins 65.3-65.7~s. Left: fit with traditionally ``Band'' function. Right: Fit to the same data with DREAM (Dissipation with Radiative Emission as a table Model) table model. These fits are based on tabulating the results of sub-photospheric energy dissipation code \cite{PW05}, and using them as input in XSPEC. See \cite{Ahlgren+15} for details.}%
  \label{fig:8}
\end{center}
\end{figure}

\def\figsubcap#1{\par\noindent\centering\footnotesize(#1)}
\begin{figure}[b]
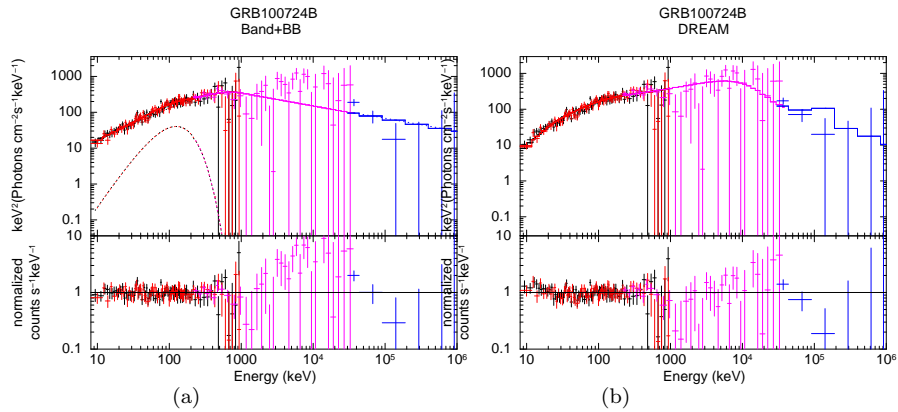
%
\begin{center}
 \parbox{2.1in}{\includegraphics[width=2in,angle=-90]{fig9a.eps}\figsubcap{a}}
  \hspace*{4pt}
\parbox{2.1in}{\includegraphics[width=2in,angle=-90]{fig9b.eps}\figsubcap{b}}
  \caption{Fits to GRB100724B at time bins 25.8-33.5~s. Left: fit with ``Band'' function plus black body. Right: Fit to the same data with DREAM model. See \cite{Ahlgren+15} for details. When fitting with a ``Band'' function, an addition of thermal component is required. However, fitting with sub-photospheric dissipation (DREAM) unifies this bursts' spectra with those of GRB090618 presented above, with the main difference being the optical depth in which the energy dissipation takes place.}%
  \label{fig:9}
\end{center}
\end{figure}

\subsubsection{Geometrical broadening}
\label{sec:4.2.2}

Even if sub-photospheric dissipation does not exist in GRB jets, still
the observed spectrum emerging from the photosphere is expected to
somewhat deviate from a pure ``Planck'' shape. The reason is the
``limb darkening'' effect: the optical path of photons emerging from
off the line of sight is larger than the optical path of photons
emerging on-axis. As a result of that, photons that originate off-axis
will be cooler than photons originating on-axis. As an observer cannot
discriminate between the two photons, the integrated spectral shape
will be a distorted ``Planck'' spectrum.

The limb darkening effect is well known in astronomy. Furthermore, the
understanding that it will play some role in shaping the observed GRB
spectra is also not new \cite{Goodman86, ANP91, Blinnikov+99}. However,
full treatment of this effect for relativistic outflows, as occur in
GRB, was carried out only recently \cite{Peer08, PR11, Begue+13,
  Aksenov+13, Ruffini+13, Vereschagin14, DZ14}.

When considering spherical, relativistic explosion characterized by
$\Gamma \gg 1$, one can show that the photospheric radius is a strong
function of the angle to the line of sight:
\beq
r_{ph} (\theta) \propto \left( {1 \over \Gamma^2} + {\theta^2 \over 3} \right)
\label{eq:r_ph_theta}
\eeq
(see Ref. \cite{Peer08}), where the proportionality constant is a
function of the mass ejection rate.

This angular dependence implies that off-axis photons are observed at
lower energies than on-axis photons, due to two effects. First, they
suffer enhanced adiabatic losses as they travel longer path below the
photosphere; and second, their Doppler boost is reduced relative to
photons emitted on-axis. Combined together, these two effects lead to
flattening of the Rayleigh-Jeans (low energy) part of the thermal
spectrum. 

\begin{figure}
\begin{center}
\includegraphics[width=10cm]{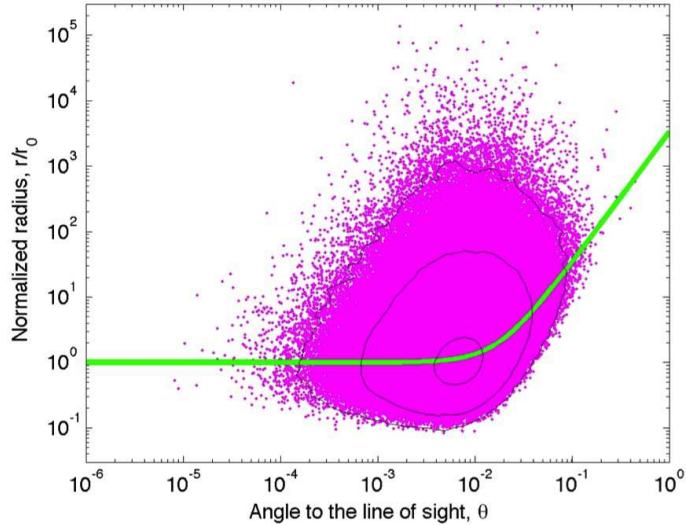}
\end{center}
\caption{The green line represent the (normalized) photospheric radius
  $r_{ph}$ as a function of the angle to the line of sight, $\theta$,
  for spherical explosion (see equation \ref{eq:r_ph_theta}). The red
  dots represent the last scattering locations of photons ejected in
  the center of relativistic expanding ``fireball'' (using a
  Monte-Carlo simulation). The black lines show contours. Clearly,
  photons can undergo their last scattering at a range of radii and
  angles, leading to the concept of ``vague photosphere''. The
  observed photospheric signal is therefore smeared both in time and
  energy.  Figure taken from \cite{Peer08}.  }
\label{fig:10}
\end{figure}

An in-depth calculation of the expected spectra, reveals the fact that
the ``photospheric radius'', defined as the surface of last
scattering, is in fact ill-defined. A photospheric radius gives only a
very crude approximation to the probability of photons to escape the
plasma (which is equal to $e^{-1}$ at $r_{ph}$). In reality, photons
have finite probability of being scattered at every location in space
where particles exist.  This realization led to the concept of a
``vague photosphere'' (See Figure \ref{fig:10}) \cite{Peer08,
  Beloborodov10, Beloborodov11, LPR13, Ruffini+13, Aksenov+13,
  Vereschagin14}. In a spherical explosion scenario, the effect of the
vague photosphere on the observed spectral shape is not large; it
somewhat modifies the Rayleigh-Jeans part of the spectrum, that reads
$F_{\nu} \propto \nu^{3/2}$ (Ref. \cite{Beloborodov11,
  DZ14}). However, this assumes an idealistic scenario of spherical
explosion, with a smooth velocity profile. More realistic numerical
models that consider outflow instabilities due to the interaction of
the jet with the stellar envelope reveal a much more pronounced effect
\cite{Lazzati+11, Ito+15}. Furthermore, as will be shortly discussed
below, for non-spherical explosion, the effect of the ``vague
photosphere'' on the observed spectrum becomes dramatic.

Even for a spherical case, emission from the ``vague photosphere''
implies that late time photons are more likely to originate from
off-axis angles. This provides a robust prediction for the late time
asymptotic decay law (assuming that the central engine is abruptly
shut), of $F(t) \propto t^{-2}$ and $T(t) \propto t^{-2/3}$
\cite{Peer08, PR11, DZ14}. This limit is obtained for the ``pure''
spherical scenario.

While the exact geometry of GRB jets, namely $\Gamma(r,\theta,\phi)$
are unknown, numerical simulations of jets propagating through the
stellar core (e.g., Ref. \cite{ZWM03}) suggest a jet profile of the form
$\Gamma(\theta) \sim \Gamma_0/(1+(\theta/\theta_j)^{2p})$, at least
for non-magnetized outflows. Such a jet profile thus assumes a
constant Lorentz factor, $\Gamma \sim \Gamma_0$ for $\theta \lsim
\theta_j$ (the ``jet core'', or inner jet), and decaying Lorentz
factor at larger angles, $\Gamma(\theta) \propto \theta^{-p}$ (outer
jet, or jet sheath). As the Lorentz factor is $\Gamma \propto L/\dot
M$, such a profile can result from excess of mass load close to the
jet edge, by mass collected from the star ($\dot M = \dot M(\theta)$),
or alternatively by angle dependent luminosity.

\begin{figure}
\includegraphics[width=8cm]{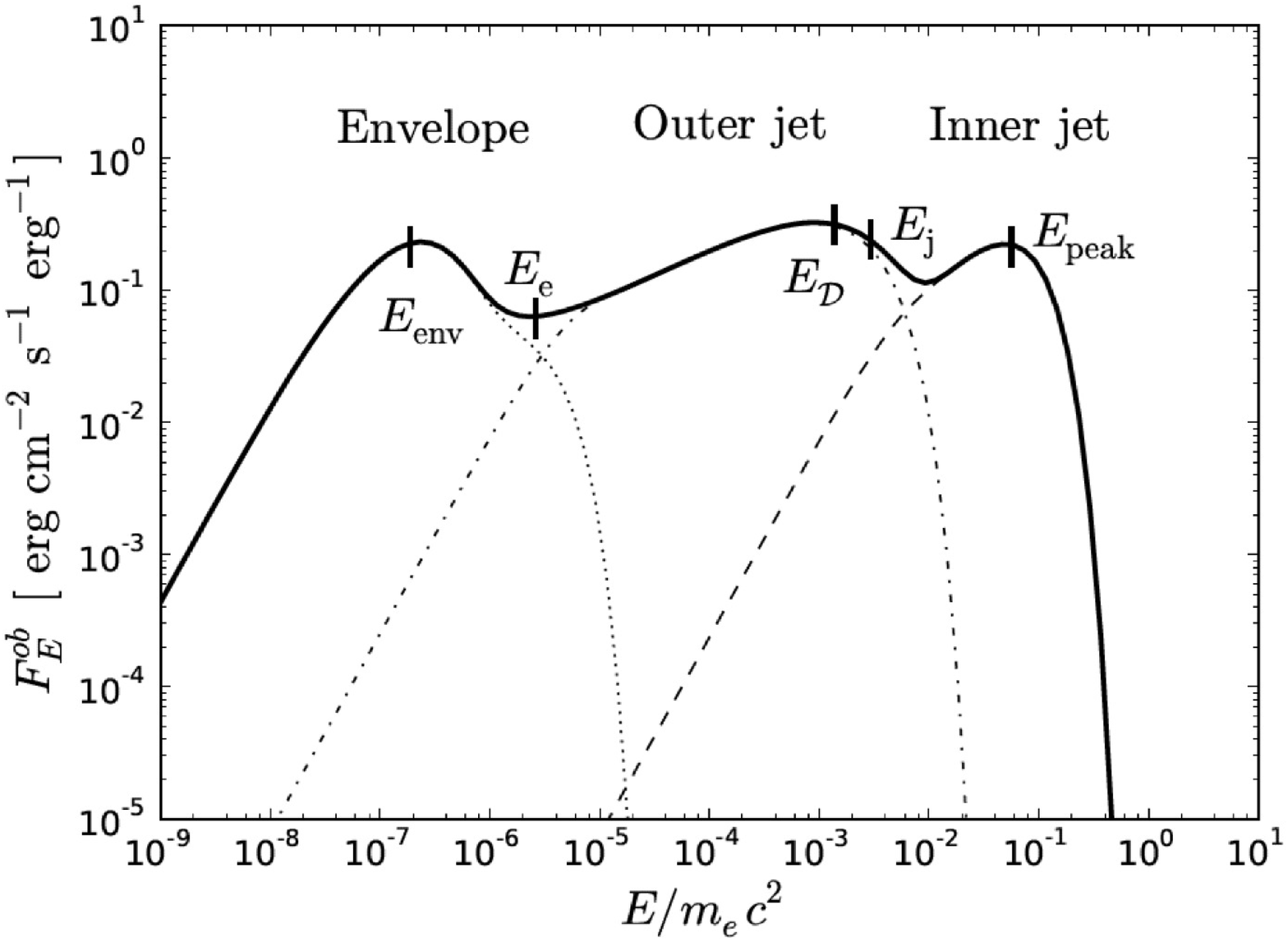}
\includegraphics[width=4cm]{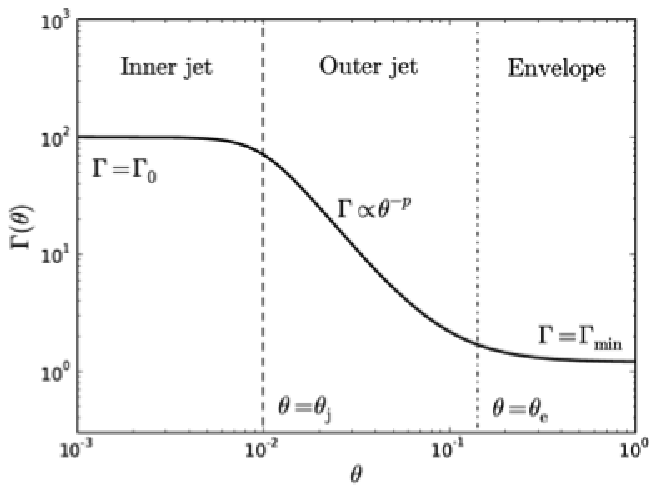}
\caption{{\bf Left.} The expected (observed) spectrum from a
  relativistic, optically thick outflow. The resulting spectra does
  not resemble the naively expected ``Planck'' spectrum. Separate
  integration of the contributions from the inner jet (where $\Gamma
  \approx \Gamma_0$), outer jet (where $\Gamma$ drops with angle) and
  envelope is shown with dashed, dot–dashed and dotted lines,
  respectively. {\bf Right.} The assumed jet profile. Figure taken
  from \cite{LPR13}.  }
\label{fig:Christoffer1}
\end{figure}

The effect of angle-dependence mass loading, $\dot M = \dot M(\theta)$
on the observed photospheric signal is dramatic.  While emission from
the inner parts of the jet result in mild modification to the black
body spectrum, photons emitted from the outer jet's photosphere
dominate the spectra at low energies (see Figure
\ref{fig:Christoffer1}, taken from Ref. \cite{LPR13}). For narrow jets
($\theta_j \Gamma_0 \lsim$~few), this leads to flat low energy
spectra, $dN/dE \propto E^{-1}$, which is independent on the viewing
angle, and very weakly dependent on the exact jet profile. This result
thus raises the possibility that the low energy slopes are in fact
part of the photospheric emission itself, even if the observed power
law is substantially different than a Rayleigh-Jeans. Furthermore, it
raises the possibility that study of the low energy slopes can be used
to infer the jet geometry.

A second aspect of this scenario is that the photospheric emission can
be observed to be highly polarized, with up to $\approx 40\%$
polarization \cite{LPR14, Chang+14, Ito+14}. While inverse-Compton
(IC) scattering produces highly polarized light, in spherical models
the polarization from different viewing angles cancels. However, this
cancellation is incomplete in jet-like models observed
off-axis. Clearly, for an off-axis observer the observed flux will be
reduced; nonetheless, for a large parameter space region it is still
high enough to be detected, in which case it will be seen to be highly
polarized \cite{LPR14}.

A non-spherical jet geometry has a third unique aspect, which is
photon energy gain by Fermi-like process. Below the photosphere,
photons are scattered back and forth between the jet core and the
sheath. Due to the difference in velocity in between the different
regions, on the average the photons gain energy. This leads to a high
energy power law tail, extending above the thermal peak \cite{LPR13,
  Ito+13}. Similar to the low energy case, this effect may potentially
be used as a new probe in studying the jet geometry [Lundman et. al.,
  in prep.].

\subsection {Implications of observations of a thermal component}
\label{sec:4.3}

A great advantage of the photospheric emission is its relative
simplicity. By definition, the photosphere is the inner most region
from which electromagnetic signal can reach the observer. Thus, the
properties of the emission site are much more constrained, relative,
e.g., to synchrotron emission (whose emission radius, magnetic field
strength and particle distribution are not known). 

This advantage enables the use of an observed thermal component as a
probe to some key parts of the underlying GRB physics. The GRB
environment is complicated, and characterized by several processes of
energy transfer that are obscured. Gravitational energy is converted
to kinetic energy (jet launching); kinetic and possibly magnetic
energy is dissipated; particles are heated; and radiation is
emitted. We can only probe the final outcome - the observed spectra
and its temporal evolution, from which all the previous stages and
their physical ingredients need to be deduced.

The relative simplicity of the thermal emission is therefore of a
great advantage, as it enables us to deduce several properties of GRB
physics that are very difficult to probe. There are four main
properties that have been discussed so far in the literature.  First,
if thermal photons are indeed the seed photons for Compton scattering,
then by comparing the thermal part to the non-thermal part of the
spectrum, one can directly probe the temperature of the hot electrons,
as well as the optical depth in which these electrons were introduced
into the plasma (which is where the energy dissipation took
place). Thus, by fitting the data, one can provide information about
the properties of the energy dissipation process. This had been
discussed in section \ref{sec:4.2.1} [see Ref. \cite{Ahlgren+15}].

Second, as discussed in section \ref{sec:4.2.2}, low and high energy
spectral slopes as well as polarization measurements may be used to
probe the geometry of GRB jets, and possibly even the viewing
angle. Nonetheless, the ability to obtain similar spectral slopes by
more than a single way implies that further theoretical work is needed
before firm conclusions could be drawn. Third, the properties of the
thermal emission could be used to infer the dynamics of the outflow;
and fourth, it may even be used to constrain the outflow
magnetization. Here, we discuss these last two probes.  A word of
caution: in order to perform these analyses one has to be able to
clearly identify the properties of the thermal component (temperature
and flux). Thus, these analysis can only be performed if the thermal
component is not strongly distorted by sub-photospheric dissipation or
geometrical effects.

\subsubsection{Probing outflow dynamics} 

In the framework of the ``hot'' fireball model in which the magnetic
field is dynamically sub-dominant, the (1-d) photospheric radius is a
function of only two parameters: the luminosity (which can be measured
once the distance is known) and the Lorentz factor. The photospheric
radius is related to the observed temperature and flux via
$r_{ph}/\Gamma \propto (F_{bb}^{ob}/\sigma {T^{ob}}^4)^{1/4}$, where
$\sigma$ is Stefan's constant, and the extra factor of $\Gamma^{-1}$
is due to light aberration.  Since $r_{ph} \propto L \Gamma^{-3}$,
measurements of the temperature and flux for bursts with known
redshift enables an independent measurement of the Lorentz factor at
the photosphere, $\Gamma$, the photospheric radius, $r_{ph}$, and the
acceleration radius, $r_0$ \cite{Peer+07}. These, in turn, can be used
to determine the full dynamical properties of the outflow.

A very interesting result is that by using this method, it is found
that $r_0$, the size of the jet base ($\Gamma(r_0) = 1$), is $r_0
\gsim 10^8$~cm, nearly two orders of magnitude above the gravitational
radius of 10~$M_\odot$ black hole \cite{Peer+07, Ryde+10, BI14,
  Peer+15} (see Figure \ref{fig:12}). While in many works it is
assumed that $r_0$ is $\approx$ few gravitational radii, in fact there
is no evidence for that in the data; the shortest variability time
scale observed in GRBs, $\delta t = r_0 /c \gsim 10$~ms, with average
value of $\approx 500$~ms \cite{GB14}. These results are therefore
consistent with the results obtained by analyzing the thermal data.

The high value of $r_0$ may be interpreted as an indication for
recollimation shocks that occur at this radius. These shocks originate
from interactions between the outflow and the collapsing star, and are
clearly seen in numerical simulations \cite{Aloy+02, ZWM03,
  Morsony+07, MA09, MI13, Lopez+13}. Thus, this result may serve as an
indirect probe for the collapsar scenario.

Furthermore, the values of the Lorentz factor found using this method
are at the range $10^2 \lesssim \Gamma \lesssim 10^3$. These values
are similar to those inferred by other methods. The results obtained
by analyzing the thermal component are aligned with recent constraints
found by Ref. \cite{VLP13}, that showed that the conditions for full
thermalization takes place only if dissipation takes place at
intermediate radii, $\sim 10^{10}$~cm, where the outflow Lorentz
factor is mild, $\Gamma \sim 10$. Interestingly, similar results
albeit with somewhat lower values of the Lorentz factor, $\Gamma \sim
10^2$ were found when analyzing X-ray flares in a similar method
\cite{Peng+14}. Thus, overall, the results obtained point towards a
new understanding of the early phases of jet dynamics.

\def\figsubcap#1{\par\noindent\centering\footnotesize(#1)}
\begin{figure}[b]%
\begin{center}
 \parbox{2.1in}{\includegraphics[width=2in]{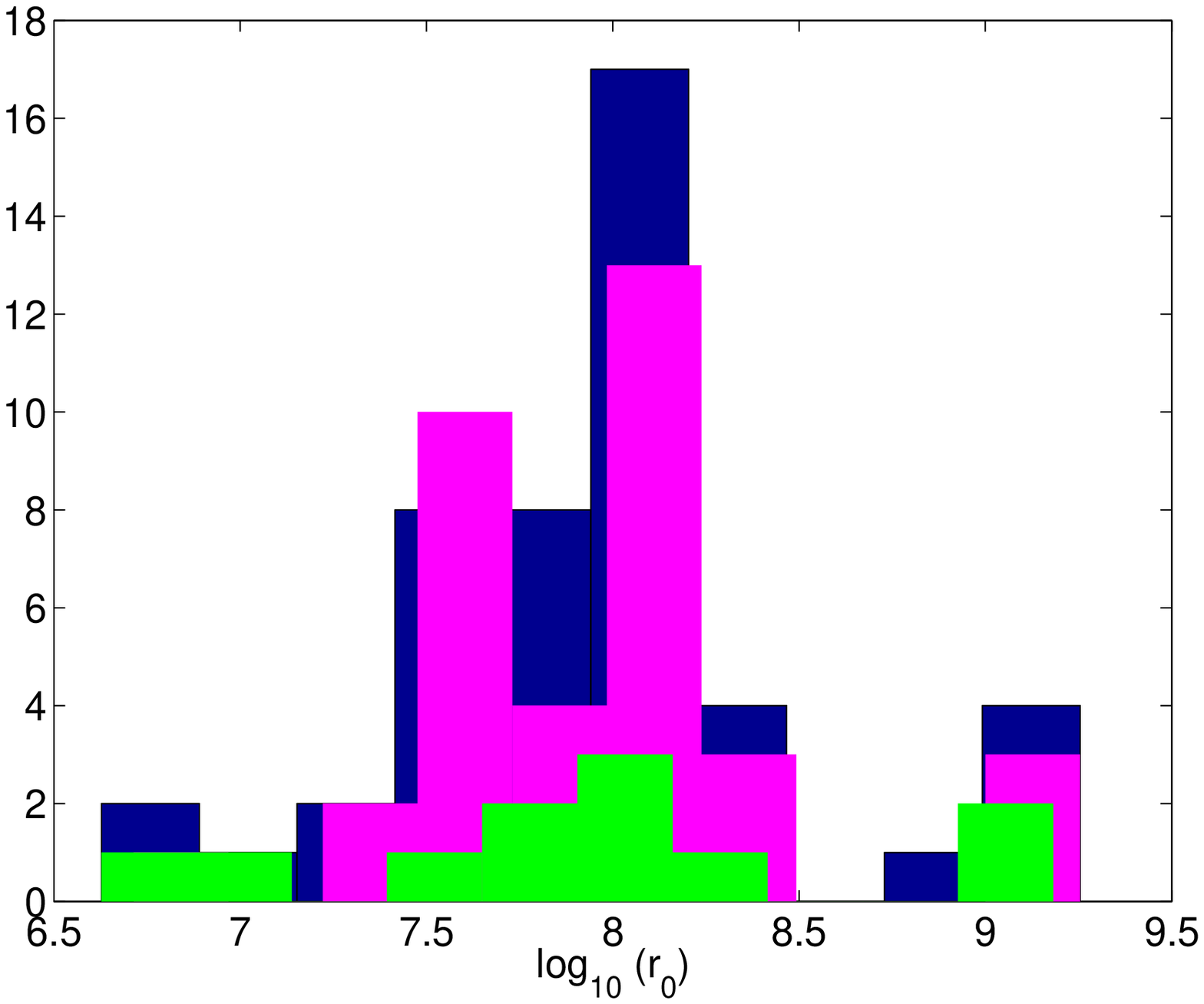}\figsubcap{a}}
  \hspace*{4pt}
 \parbox{2.1in}{\includegraphics[width=2in]{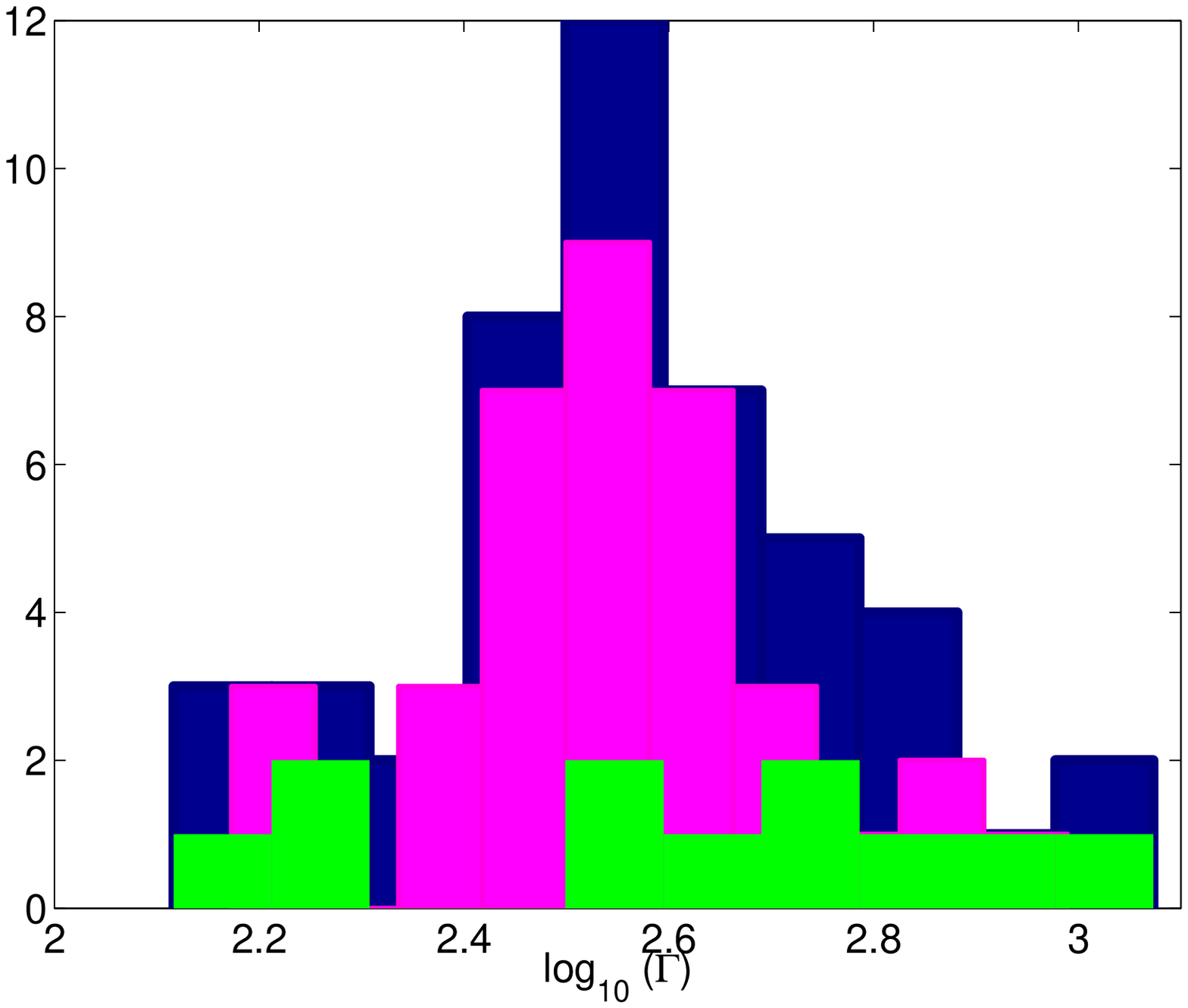}\figsubcap{b}}
  \caption{Histograms of the mean values of $\log_{10} (r_0)$ (left) and
  $\log_{10} (\Gamma)$ (right) deduced from analyzed the properties of the thermal component in 47 GRBs. Blue are for the entire sample, while
  magenta are for 36 GRBs in category (III)
  sample only (which is a homogeneous sub-sample), and green are for 11 GRBs in categories (I) and (II). See Ref. \cite{Peer+15} for details.}%
  \label{fig:12}
\end{center}
\end{figure}

\subsubsection{Probing outflow magnetization}

One of the key open questions in the study of GRBs is the role played
by the magnetic fields. Within the framework of the original
``fireball'' model \cite{RM92, RM94}, the flow accelerates to
relativistic velocities by radiative pressure, and magnetic fields are
dynamically unimportant. They do, though, play an important role in
extracting the energy from the hot electrons that radiate via
synchrotron mechanism.

In contrast to this picture, the leading mechanism for accelerating
jets in active galactic nuclei (AGNs) is the Blandford-Znajek process
\cite{BZ77}, which involves strong magnetic fields. It was therefore
suggested that magnetic fields may be energetically dominated, hence
play a central role in determining the dynamics of GRB as well
\cite{SDD01, LK01, Drenkhahn02, DS02}.  This scenario could be valid
if the progenitor of GRBs are rapidly spinning, strongly magnetized
neutron star - the so called ``magnetars'' \cite{Usov92, Thompson94,
  DL98, Wheeler+00, Metzger+08, Bucciantini+09, Metzger+11,
  Bucciantini+09, Bucciantini+12}.  In this case, the main source of
energy available for heating the particles is reconnection of the
magnetic field lines \cite{Thompson94, SDD01, Drenkhahn02, DS02},
possibly enhanced by turbulentic outflows \cite{ZY11}.

\begin{figure}
\begin{center}
\includegraphics[width=8cm]{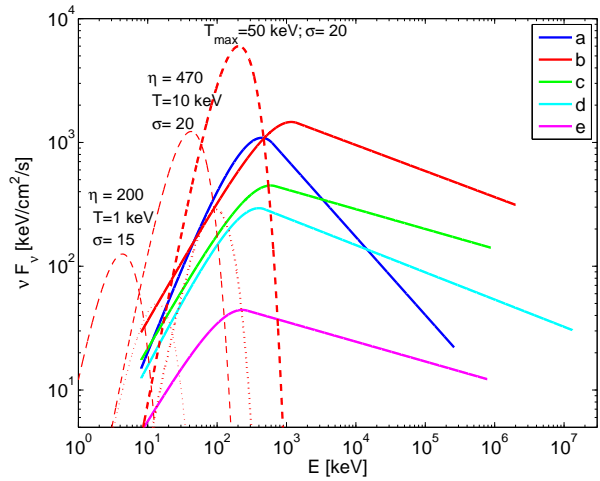}
\end{center}
\caption{Observed Band-function spectra for the five epochs of
  GRB080916C, taken from Ref. \cite{Abdo+09b} (color solid) and the
  predicted lower limits of the photosphere spectra (red dashed) for
  different parameters for the epoch (b) within the framework of the
  baryonic fireball models. Red, thick-dashed curve: the internal
  shock model with $\delta t^{\rm ob} = 0.5$ s, corresponding to
  $T^{\rm ob}_{ph} = 50$~keV; red, thin-dashed curves: for $T^{\rm
    ob}_{ph} = 10, 1$~keV. The suppressed photosphere spectra are
  plotted by red, dotted curves, with the required σ values
  marked. Figure taken from \cite{ZP09}.  }
\label{fig:13}
\end{figure}

There are several differences between magnetically dominated outflow
and baryonic dominated outflows. One such difference is the location
of the photospheric radius, which has a somewhat different dependence
on the free model parameters. A second difference is the fact that the
strong magnetic fields serve as ``energy reservoir'', dissipating
their energy gradually. This implies that the flux of the thermal
photons is weaker in magnetized models in comparison with
baryon-dominated ones. Based on this realization, it was argued that a
weak - or lack thereof of a thermal component could be attributed to a
strong outflow magnetization \cite{ZM02, DM02, ZP09}. This argument
was used by Ref. \cite{ZP09} to claim that the outflow in GRB080916,
which did not show any clear evidence for the existence of a thermal
component, could be highly magnetized, with $\sigma \geq 20$ (see
Figure \ref{fig:13}). This model further predicts high polarization
\cite{Deng+16}.

Furthermore, strong magnetic field would lead to rapid radiative
cooling of the energetic particles. This puts strong constraints on
the properties of the particle acceleration mechanism that could
reproduce the observed signal \cite{BP14}. In a recent work
\cite{BP15}, it was shown that in the framework of continuous magnetic
reconnection model, conditions for full thermalization do not exist in
the entire region below the photosphere. As a result, the produced
photons are up-scattered, and the resulting peak of the Wien
distribution formed is at $\gsim 10$~MeV. This again leads to the same
conclusion as drawn above, namely that identification of thermal
component at energies of $\lsim 100$~keV must imply that the outflow
cannot be highly magnetized.

\section{Summary}
\label{sec:5}

A major breakthrough in our understanding of GRB prompt emission
occurred in recent years, with the realization that a thermal emission
component exists on top of the over all non-thermal spectra. This
realization opens up a completely new window into studying the physics
of GRBs. In this short review, we highlighted some of the major
aspects of this realization. 

In section \ref{sec:2} we pointed out to the fact that thermal
emission was predicted already by the very early models of
cosmological GRBs. It was later abandoned, as the observed spectra did
not reveal a clear evidence of a ``Planck'' spectrum. However, it was
re-considered in the early 2000's, following the realization that
known non-thermal models suffer difficulties in fitting the observed
data. 

In section \ref{sec:3}, we described the observational status. There
are several key results that need to be emphasized.
\begin{enumerate}
\item The ``Band'' function provides good fits to most of GRB data, with
  only a relatively small fraction of GRBs that are of an
  exception. Nonetheless, the use of ``Band'' fits is highly
  misleading, as the ``Band'' model, from its very nature, is not
  capable of capturing any ``wiggles'' that may indicate the
  existence of a thermal emission.  Furthermore, by definition, most
  bursts are detected close to the detection limit, in which case a
  weak thermal signal could not be observed.
\item 
The fraction of GRBs in which a thermal component is detected
increases with their observed luminosity. In most cases in which a
thermal component was detected, it was accompanied by a non-thermal
emission. Furthermore, attempts to associate a sole non-thermal
radiative mechanism to the observed spectra show inconsistency. These
facts suggest that a thermal component is in fact very ubiquitous
among GRBs.
\item 
In all cases in which a thermal component was detected, both the
temperature and thermal flux show well defined, repetitive temporal
behavior, which is distinct from the non-thermal behavior. Although a
theory that can explain this behavior is of yet incomplete, the
repetitive behavior strengthen the interpretation of this component
to be distinct.
\item 
As a consequence, in order to make further progress, the logical step
is to abandon the ``Band'' fits, and fit the data with
physically-motivated models, that would include a thermal component,
in addition to non-thermal emission processes. Several such models
already exist, though they are still not in wide use. We can
anticipate that with a more wider use, the existence of a thermal
emission would become more and more clear.
\end{enumerate}

Section \ref{sec:4} was devoted to an overview of the theoretical
status.  We pointed out that all leading theoretical models predict
the existence of a thermal component, though no existing theory
provides robust predictions about its strength.  We then discussed
various mechanisms that act to broaden the naively expected ``Planck''
function. As we showed, the ``Planck'' function may be so heavily
distorted, that the resulting spectra would resemble the observed
one. If this is indeed the case, then the thermal component plays a
very central role in the entire observed prompt emission. In
particular, we discussed the following points:
\begin{enumerate}
\item Sub-photospheric energy dissipation is expected by many
  theoretical models. If the dissipation occurs not too-far below the
  photosphere, a ``two temperature'' plasma emerges. In this case,
  there is a complicated connection between the thermal and
  non-thermal parts of the spectrum, as the thermal photons serve as
  seed photons for scattering by the hotter electrons. In this
  scenario, the leading radiative process above the thermal peak is IC
  scattering, rather than synchrotron.
\item Relativistic ``limb darkening'' effect will further broaden the
  ``Planck'' spectra, irrespective of any energy dissipation that may
  or may not exist. Study of this effect lead to the realization that
  the photosphere is, in fact ``vague''. While this results in only a
  minor modification in the spherical explosion case, it has a
  dramatic effect on the observed spectra if the outflow is not
  spherical. In this later case, photons can be accelerated by
  Fermi-like process below the photosphere.
\item
If the thermal emission is not strongly distorted, its properties can
be used as a direct probe of the dynamics of the outflow. In
particular, it can provide an indirect evidence for the
``collapsar'' model. If the outflow is highly magnetized, the thermal
component is expected to weaken. Therefore, weak, or lack of thermal
component can be used to constrain the outflow magnetization.
\end{enumerate}

Nearly all of the realizations described here- both observational and
theoretical - occurred only in the last decade or so. Thus, while a
major progress had been made in recent years, clearly there are still several very
important open questions in the study of GRBs. These include, e.g.,
the questions of progenitor, magnetization and energy dissipation. 

It is difficult to state at this point the role that thermal emission
will play in the future in resolving these issues. A main concern is
the fact that the observed signal is often degenerated, namely it can
be explained by more than one model. A good example is the fact that a
weak ``Planck'' component can result from either (1) adiabatic losses;
(2) strong distortion due to sub-photospheric dissipation; or (3)
strong magnetization. Each of these models is very different in nature
than the other ones. Thus, one needs to combine the thermal signal
with additional clues - both observational (broad-band non-thermal
signal, temporal evolution) as well as theoretical models, in order to
achieve a comprehensive understanding of GRB physics. Nonetheless, we
believe that it is clear that the study of a thermal component will
continue to provide new probes that will eventually lead to answering
the open questions.

\section*{Acknowledgments}
We would like to thank Bing Zhang for many useful comments.
AP wishes to acknowledge support from the European Union Seventh Framework Programme
(FP7/2007-2013) under grant agreement ${\rm n}^\circ$ 618499.

\bibliographystyle{ws-procs975x65}

\end{document}